\documentclass[12pt]{article}
\usepackage{epsfig,amsfonts,amssymb}
\usepackage{cite}
\input epsf.sty
\def\ZZZ{{\hbox{ Z\kern-1.6mm Z}}}
\def\RRR{{\hbox{ R\kern-2.4mm R}}}
\def\CCC{{\hbox{ C\kern-2.0mm C}}}
\def\zzz{{\hbox{z\kern-1mm z}}}

\newcommand{\qq}{k}
\newcommand{\pp}{l}

\newcommand{\qeq}{{\hbox{=\kern-2.3mm ? \kern.5mm }}}
\renewcommand{\qeq}{=}

\newcommand{\BB}{{\cal B}}

\newcommand{\AAA}{{\cal A}}

\newcommand{\FF}{{\cal F}}

\newcommand{\wt}{\widetilde}
\newcommand{\wh}{\widehat}

\newcommand{\NN}{{\cal N}}

\newcommand{\be}{\begin{equation}}
\newcommand{\ee}{\end{equation}}
\newcommand{\ben}{\begin{eqnarray}\displaystyle}
\newcommand{\een}{\end{eqnarray}}

\newcommand{\bea}[1]{\begin{eqnarray}\label{#1} }
\newcommand{\eea}{\end{eqnarray}}

\newcommand{\refb}[1]{(\ref{#1})}
\newcommand{\p}{\partial}
\newcommand{\sectiono}[1]{\section{#1}\setcounter{equation}{0}}

\def\one{{\hbox{ 1\kern-.8mm l}}}
\def\zero{{\hbox{ 0\kern-1.5mm 0}}}

\begin{document}

\begin{center}
{\Large \bf
Adding Charges to N=4 Dyons}

\end{center}

\vskip .6cm
\medskip

\vspace*{4.0ex}

\centerline{\large \rm  }

\vspace*{4.0ex}

\centerline{\large \rm Nabamita Banerjee,
Dileep P. Jatkar and Ashoke Sen}

\vspace*{4.0ex}

\centerline{\large \it Harish-Chandra Research Institute}

\centerline{\large \it  Chhatnag Road, Jhusi,
Allahabad 211019, INDIA}

\vspace*{1.0ex}
\centerline{E-mail:  nabamita, dileep, sen@mri.ernet.in}

\vspace*{5.0ex}

\centerline{\bf Abstract} \bigskip

The spectrum of dyons in a class of N=4 supersymmetric string
theories has been
found for a specific set of electric and magnetic charge vectors.
We extend the analysis to more general charge vectors by
considering various charge carrying collective excitations of the
original system.

\vfill \eject

\baselineskip=18pt

\tableofcontents

\sectiono{Introduction} \label{s1} 

We now have a good understanding
of the spectrum of quarter BPS dyons in a class of $\NN=4$
supersymmetric string 
theories\cite{9607026,0505094,0506249,0510147,0602254,
0603066,0605210,0607155,0609109,0612011,0702141,0702150},  
obtained
by taking a $\ZZZ_N$ orbifold of type II string theory compactified
on $K3\times T^2$ or $T^6$. However, in the
direct approach to the computation of the spectrum based on
counting of states the spectrum has so far been computed only for
states carrying a restricted set of 
charges\cite{0605210,0607155,0609109}. 
Our goal in this
paper will be to extend this analysis to states carrying a more general
set of charges, obtained from collective excitations of the system
that has been analyzed earlier. 
For simplicity we shall restrict our analysis to 
type II string theory compactified on  $K3\times T^2$.
Generalizing this to the case of $\NN=4$ supersymmetric orbifolds
of this theory is straightforward, requiring setting to zero
some of the 
charges 
which
are not invariant under the orbifold group. The analysis
for $\NN=4$ supersymmetric $\ZZZ_N$ orbifolds of type II string
theory compactified on $T^6$ can also be done in an identical manner.

\sectiono{Background} \label{s2}

We consider the case of type IIB string theory
on $K3\times S^1\times \wt S^1$
or equivalently heterotic string theory
on $T^4\times S^1\times \wh S^1$. The latter
description -- to be called the second description -- is obtained from the
first description by first making an S-duality 
transformation in ten dimensional type IIB string theory, followed
by a T-duality along the circle $\wt S^1$ that converts it to type
IIA string theory on $K3\times S^1\times \wh S^1$ and then using the
six dimensional string-string duality that converts it to heterotic string
theory on $T^4\times S^1\times \wh S^1$. 
The coordinates 
$\psi$, $y$ and $\chi$ along
$\wt S^1$, $S^1$ and $\wh S^1$ are all normalized to have period
$2\pi\sqrt{\alpha'}$. Other
normalization and sign
conventions have been described in appendix \ref{sa}.

The compactified theory
has 28 U(1) gauge fields and hence a given state
is characterized by 28 dimensional electric and
magnetic charge vectors $Q$ and $P$. 
We shall use the second 
description
of the theory to classify charges as electric and magnetic. Since in
this description there are no RR fields or D-branes, 
an electrically charged state
will correspond to an elementary string state and a magnetically
charged state will correspond to wrapped 
NS 5-branes and Kaluza-Klein
monopoles. The relationship between $Q$ and $P$ and the fields
which appear in the supergravity theory underlying the second
description will follow
the convention of \cite{0508042} in the $\alpha'=16$
unit. This has been reviewed in appendix \ref{sa}, 
eq.\refb{et.5}.
In this description the theory has an $SO(6,22;\ZZZ)$ T-duality
symmetry, and 
the T-duality 
invariant combination of
charges is given by 
\be \label{e0}
Q^2 = Q^T L Q, \qquad P^2 = P^T L P, \qquad Q\cdot P
= Q^T L P\, ,
\ee
where $L$ is a symmetric matrix with 22 eigenvalues $-1$ and $6$
eigenvalues $+1$.
We shall choose a basis in which $L$  has the form
\be \label{e1}
L =  \pmatrix{\wh L & & & & \cr & 0 & 1 & & \cr & 1 & 0 &&\cr
&&& 0_2 & I_2\cr &&& I_2 & 0_2}\, ,
\ee
where $\wh L$ is a matrix with 3 eigenvalues $+1$ and 19
eigenvalues $-1$. The charge vectors
will be labelled as
\be \label{e1.1}
Q = \pmatrix{\wh Q\cr \qq_1 \cr \qq_2 \cr \qq_3 \cr \qq_4
\cr
\qq_5\cr \qq_6}\, , \qquad P = \pmatrix{\wh P\cr \pp_1\cr \pp_2 \cr \pp_3
\cr \pp_4\cr \pp_5 \cr \pp_6}
\, .
\ee

According to the convention of appendix \ref{sa},
$\qq_3$, $\qq_4$, $-\qq_5$ and
$-\qq_6$ label respectively the momenta along $\wh S^1$, $S^1$
and fundamental string winding along $\wh S^1$ and $S^1$
in the second description of the theory. On
the other hand $-\pp_3$, $\pp_4$, $\pp_5$ and $\pp_6$ label 
respectively the number of
NS 5-branes wrapped along $S^1\times T^4$ and $\wh S^1\times T^4$, 
and Kaluza-Klein
monopole charges associated with $\wh S^1$ and $S^1$.
The rest of the charges label the momentum/winding
or monopole charges associated with the other internal directions. 
By following the duality chain that relates the first and second
description of the theory and using the sign convention of appendix
\ref{sa},
the different components of $P$ and $Q$
can be given the following interpretation in
the first description of the theory.
$\qq_3$ represents 
the D-string winding charge along $\wt S^1$, $\qq_4$
is the momentum along $S^1$, $\qq_5$ is the D5-brane charge
along $K3\times \wt S^1$, $\qq_6$ is the number of Kaluza-Klein
monopoles associated with the compact circle $\wt S^1$,
$\pp_3$ is the D-string winding charge along $ S^1$, $-\pp_4$
is the momentum along $\wt S^1$, $\pp_5$ is the D5-brane charge
along $K3\times   S^1$ and $\pp_6$ is the number of Kaluza-Klein
monopoles associated with the compact circle $ S^1$.
Other components of $Q$ ($P$) represent various other branes of
type IIB string theory wrapped on $\wt S^1$ ($S^1$) times various
cycles of $K3$. We shall choose a convention in which
the 22-dimensional 
charge vector $\wh Q$ represents 
3-branes wrapped along the 22 2-cycles of K3 times $\tilde S^1$,
$\qq_1$ represents fundamental type IIB string winding charge
along $\wt S^1$, $\qq_2$ represents the number of
NS 5-branes of type IIB wrapped along $K3\times \wt S^1$, 
 the 
22-dimensional 
charge vector $\wh P$ represents 
3-branes wrapped along the 22 2-cycles of K3 times $ S^1$,
$\pp_1$ represents fundamental type IIB string winding charge
along $S^1$ and $\pp_2$ represents the number of
NS 5-branes of type IIB wrapped along $K3\times  S^1$.
In this convention
$\wh L$ represents the intersection matrix of 2-cycles of $K3$.
Using the various sign conventions described in appendix \ref{sa}, 
and the
T-duality transformation laws for the RR fields given in \cite{9910053}
one can verify that the combinations $\qq_3 \qq_5 + \wh Q^2/2$,
$\pp_3 \pp_5 + \wh P^2 /2$ and $\qq_3 \pp_5+\pp_3 \qq_5+\wh Q\cdot \wh P$
are invariant under the mirror symmetry
transformation on $K3$.

The original configuration studied in \cite{0605210}
has charge vectors of the form:
\be \label{e1.2}
Q = \pmatrix{\wh 0\cr 0 \cr 0 \cr 0 \cr -n
\cr
0\cr -1}\, , \qquad P = \pmatrix{\wh 0\cr 0\cr 0 \cr Q_1
-Q_5=Q_1-1 \cr -J  \cr Q_5=1 \cr 0}
\, .
\ee
Thus in the first description
we have $-n$ units of momentum along $S^1$, $J$ units of
momentum along $\wt S^1$, a
single Kaluza-Klein monopole (with negative magnetic charge)
associated 
with $\wt S^1$, a single
D5-brane wrapped on $K3\times S^1$ and $Q_1$ D1-branes wrapped
on $S^1$. The D5-brane wrapped on $K3\times  S^1$ also
carries $-1$ units of D1-brane charge along $S^1$; this is responsible
for the shift by $-1$ of $Q_1$ as given in \refb{e1.2}.
The associated invariants are
\be \label{e1.3}
Q^2 = 2n, \qquad P^2 = 2(Q_1-1), \qquad Q\cdot P = J\, .
\ee
The degeneracy of this system was calculated in \cite{0605210}
as a function of $n$, $Q_1$ and $J$. If we call this function
$f(n, Q_1, J)$, then we can express the degeneracy $d(Q,P)$ as a
function of $Q,P$ as:
\be \label{e1.4}
d(Q,P) = f\left( {1\over 2} Q^2, {1\over 2}P^2+1, Q\cdot P\right)\, .
\ee
Ref.\cite{0605210} actually
considered a more general charge vector where $Q_5$, representing the
number of D5-branes wrapped along $K3\times S^1$, was arbitrary and
derived the same formula \refb{e1.4} for $d(Q,P)$.
However, the analysis of dyon spectrum becomes simpler for
$Q_5=1$. For this reason we have set $Q_5=1$. We shall comment
on the 
more general
case at the end.

\sectiono{Charge Carrying Deformations} \label{s3}

Our goal will be to consider charge vectors more general than
the ones given in \refb{e1.2} and check if the degeneracy is
still given by \refb{e1.4}. We shall do this by adding charges to the
existing system by exciting appropriate collective modes of the
system. These collective modes come from three sources:
\begin{enumerate}
\item The original configuration in the type IIB theory contains
a Kaluza-Klein monopole associated with the circle $\wt S^1$.
This solution is given by 
\be\label{tnutgeom}
ds^2 = \left(1+\frac{K\sqrt{\alpha'}}{2r}\right) 
\left( dr^2 +  r^2 ( d\theta^2 + \sin^2 \theta d\phi^2) \right)
+ K^2\left( 1 + \frac{K\sqrt{\alpha'}}{2r}
\right)^{-1} \left( \, d\psi + {\sqrt{\alpha'}\over 2}\cos\theta d\phi
\right)^2
\ee
with the identifications:
\be\label{eident}
(\theta,\phi,\psi) \equiv (2\pi -\theta,\phi+\pi, \psi+{\pi\over 2}
\sqrt{\alpha'})
\equiv (\theta,\phi+2\pi,\psi+\pi\sqrt{\alpha'})\equiv 
(\theta,\phi,\psi+2\pi\sqrt{\alpha'})\, .
\ee
The coordinate $\psi$ can be regarded as the coordinate of $\wt S^1$,
whereas $(r,\theta,\phi)$ represent spherical polar coordinates of the
non-compact space.
$K$ is a constant related to the physical radius of $\wt S^1$.
This geometry, also known as the Taub-NUT
space, admits a normalizable self-dual harmonic form $\omega$, given
by\cite{brill,pope}
\be \label{eomegadef}
\omega \propto {2\over \sqrt{\alpha'}}
{r\over r +{1\over 2} K\sqrt{\alpha'}} 
d\sigma_3 + {K\over (r+{1\over 2}K\sqrt{\alpha'})^2}dr
\wedge \sigma_3\, , \qquad \sigma_3 \equiv
\left (d\psi + {\sqrt{\alpha'}\over 2} \cos\theta d\phi\right)\, .
\ee
\refb{tnutgeom} represents the geometry of the space-time transverse
to the Kaluza-Klein monopole. 
Besides the $K3$ surface, the world-volume of the Kaluza-Klein
monopole spans the circle $S^1$, which we shall label 
by $y$, and time $t$.

Now type IIB string theory compactified on K3 has various 2-form fields,
-- the original NSNS and RR 2-form fields $B$ and $C^{(2)}$
of the ten dimensional
type IIB string theory 
 as well as the components of the 4-form field $C^{(4)}$ along
various 2-cycles of $K3$.
Given any such 2-form field $C_{MN}$, 
we can introduce a scalar
mode $\varphi$ by considering deformations of the form\cite{9705212}:
\be \label{e1.5}
C = \varphi(y,t) \, \omega\, ,
\ee
where $y$ denotes the coordinate along $S^1$. If the field strength
$dC$ associated with $C$ is
self-dual or anti-self-dual in six dimensions then the corresponding
scalar field $\varphi$ is chiral in the $y-t$ space; otherwise it represents
a non-chiral scalar field. 
We can now consider configurations which carry momentum conjugate
to this scalar field $\varphi$ or
winding number along $y$
of this scalar field $\varphi$,  represented
by a solution where $\varphi$ is linear in $t$ or $y$. In the six
dimensional language this corresponds to $dC \propto dt\wedge \omega$
or
$dy\wedge\omega$. {}From \refb{eomegadef} we see that
$dC\propto  dt\wedge \omega$
will have a component proportional to $r^{-2}\, 
dt\wedge dr\wedge d\psi$ for large $r$, and hence 
the coefficient of this term represents the charge associated with
a string, electrically charged under $C$, wrapped 
along $\wt S^1$. On the
other hand $dC\propto  dy\wedge \omega$
will have a component proportional to $\sin\theta\,
dy\wedge d\theta\wedge d\phi$
and the coefficient of this term will represent 
the charge associated with
a string, magnetically charged under $C$, wrapped 
along $\wt S^1$. If the 2-form field $C$ represents the original
RR or
NSNS 2-form field of type IIB string theory in ten dimensions,
then the electrically charged string would correspond to a D-string or
a fundamental type IIB string and the magnetically charged string
would correspond to a D5-brane or NS 5-brane wrapped on $K3$.
On the other hand if the 2-form $C$ represents the
component of the 4-form field along a 2-cycle of K3, then the
corresponding string represents a D3-brane wrapped on a
2-cycle times $\wt S^1$. Recalling the interpretation
of the charges $\wh Q$ and $\qq_i$ appearing in \refb{e1.1} we now
see that
the momentum and winding modes of
$\varphi$ correspond to the charges $\wh Q$, 
$\qq_1$, $\qq_2$, $\qq_3$ and
$\qq_5$. More specifically, after taking into account the sign
conventions described in appendix \ref{sa}, 
these charges correspond to switching on
deformations of the form:
\ben \label{ep1}
&& dB \propto -\qq_1 dt\wedge \omega,  \quad
dB\propto \qq_2 dy\wedge \omega, \quad 
d C^{(2)} \propto -\qq_3 dt\wedge \omega,  
\quad dC^{(2)}\propto \qq_5 dy\wedge \omega,
\nonumber \\ 
&& dC^{(4)} \propto \sum_\alpha \wh 
Q_\alpha (1 + *) \, \Omega_\alpha
\wedge dy\wedge \omega \, ,
\een
where $\{\Omega_\alpha\}$ denote a basis of harmonic 2-forms
on $K3$ ($1\le\alpha\le 22$) satisfying
$\int_{K3}\Omega_\alpha\wedge\Omega_\beta=\wh L_{\alpha\beta}$.
Thus in the presence of these deformations 
we have a more general electric charge vector of the form
\be \label{e1.6}
Q_0 = \pmatrix{\wh Q\cr \qq_1 \cr \qq_2 \cr \qq_3 \cr -n
\cr
\qq_5\cr -1}\, .
\ee

As can be easily seen from \refb{ep1}, 
$\qq_2$ represents NS 5-brane charge wrapped along 
$K3\times \wt S^1$.  However, for weakly coupled type IIB string theory,
the presence of this charge could have large backreaction on the
geometry. In order to avoid it we shall choose
\be \label{e1.7}
\qq_2 = 0\, .
\ee

\item The original configuration considered in \cite{0605210} also
contains a D5-brane wrapped around $K3\times S^1$. We can
switch on flux of world-volume gauge field strengths $\FF$
on the D5-brane
along the various 2-cycles of K3 that it wraps. 
The net coupling of the RR gauge fields to the D5-brane in
the presence of the world-volume gauge fields may be expressed 
as\cite{9910053}
\be \label{etot}
\int \left[ C^{(6)} + C^{(4)}\wedge \FF + {1\over 2} C^{(2)}\wedge \FF
\wedge \FF + \cdots\right]\, ,
\ee
up to a constant of proportionality. The integral is over the D5-brane
world-volume spanned by $y$, $t$ and the coordinates of $K3$.
In order to be compatible with the convention of appendix \ref{sa}
that the 
D5-brane
wrapped on $K3\times S^1$ carries negative 
$(dC^{(6)})_{(K3)yrt}$ asymptotically, we need to take the
integration measure in the $yt$ plane in \refb{etot}
as $dy\wedge dt$, \i.e.\ $\epsilon^{yt}>0$.
Via the coupling
\be \label{e1.8}
\int C^{(4)} \wedge \FF \, ,
\ee
the gauge field
configuration will produce the charges of a D3-brane wrapped on
a 2-cycle of $K3$ times $S^1$, -- \i.e.\ the 22 dimensional 
magnetic charge vector $\wh P$ appearing in \refb{e1.1}.
More precisely, we find that
the gauge field flux required to produce a 
specific
magnetic charge vector $\wh P$ is
\be \label{eflux1}
\FF\propto -\sum_\alpha \wh P_\alpha \, \Omega_\alpha\, .
\ee

\item The D5-brane can also carry
electric flux along $S^1$. This will induce the charge of a fundamental
type IIB string wrapped along $S^1$. According to the physical
interpretation of various charges given earlier,
this gives the component $\pp_1$ of the magnetic charge vector $P$.

The net result of switching on both the electric and magnetic flux
along the D5-brane world-volume is to generate a magnetic charge
vector of the form:
\be \label{emag}
P_0 = \pmatrix{\wh P\cr \pp_1\cr 0 \cr Q_1
-1 \cr -J  \cr 1 \cr 0}
\, .
\ee

\end{enumerate}

\sectiono{Additional Shifts in the Charges} \label{s4}
This, however, is not the end of the story. So far we have
discussed the effect of the
various collective mode excitations on the charge vector to first
order in the charges.  We have not taken into account the effect 
of the interaction of deformations produced by the collective
modes with the background fields already present in the system,
or the background fields produced by other collective modes.
Taking into account these effects produces further shifts in the
charge vector as described below.
\begin{enumerate}
\item As seen from \refb{etot}, 
the D5-brane world-volume theory has a coupling
proportional to
$\int C^{(2)}\wedge \FF\wedge \FF$.
Thus in the presence of magnetic flux
$\FF$ the D5-brane wrapped on $K3\times S^1$ acts as a source of the
D1-brane charge wrapped on $S^1$. The effect is a shift in the magnetic
charge quantum number $\pp_3$ that is quadratic 
in $\FF$ and hence 
quadratic in $\wh P$ due to
\refb{eflux1}. A careful calculation, taking into
account 
various signs and normalization factors,  shows that the
net effect of this term is to give an additional contribution to $\pp_3$ 
of the form:
\be \label{emag2}
\Delta_1 \pp_3 =  -\wh P^2 /2  
\, .
\ee
\item Let $C$ be a 2-form in the six dimensional theory obtained by
compactifying type IIB string theory on $K3$ and $F=dC$ be
its field strength. As summarized in \refb{ep1}, switching on various
components of the electric charge vector $Q$ requires us to switch
on $F$ proportional to $
dt\wedge\omega$ or $ dy\wedge\omega$.
The presence  of such background induces a coupling 
proportional to
\be \label{eq1}
-\int \sqrt{-\det g} g^{yt} F_{ymn} F_{t}^{~mn} 
\ee
with the indices $m,n$ running over the coordinates 
of the Taub-NUT space. This
produces a source for $g^{yt}$, \i.e.\ momentum along $S^1$. 
The effect of such terms is to shift the component $\qq_4$ of the charge
vector $Q$. A careful calculation shows that the net change in
$\qq_4$ induced due to this coupling is given by
\be \label{eq2.1}
\Delta_2 \qq_4 =  \qq_3 \qq_5 + \wh Q^2/2
\, ,
\ee
where we have used the fact that $\qq_2$ has been set to zero.
The $\qq_3\qq_5$ term comes from taking $F$ if \refb{eq1} to be the
field strength of the RR 2-form field, and $\wh Q^2/2$ term
comes from taking $F$ to be the
field strength of the components of the RR 4-form field along various
2-cycles of $K3$.
\item The D5-brane wrapped on $K3\times S^1$
or the magnetic flux on this brane along any of the 2-cycles of $K3$
produces a magnetic type 2-form field configuration of the form:
\be \label{eq2}
F\equiv dC \propto \sin\theta\, d\psi \wedge d\theta\wedge d\phi\, ,
\ee
where $C$ is any of the RR 2-form fields in six dimensional
theory obtained by compactifying type IIB string theory on $K3$.
One can verify that the 3-form appearing on the right hand side
of \refb{eq2} is both closed and co-closed in the Taub-NUT
background and hence $F$ given in \refb{eq2} satisfies both the 
Bianchi identity and the linearized equations of motion.
The coefficients of the term given in \refb{eq2} for various
2-form fields $C$ are determined in terms of $\wh P$ and the
D5-brane charge along $K3\times S^1$ which has been set equal
to 1.
This together with the term in $F$ proportional to 
$dt\wedge \omega$ coming from the collective coordinate
excitation of the Kaluza-Klein monopole  generates a source
of the component $g^{\psi t}$ of the metric via the coupling
proportional to
\be \label{eq3}
-\int \sqrt{-\det g} g^{\psi t} F_{\psi mn} F_{t}^{~mn} 
\ee
This induces a net momentum along $\wt S^1$ and gives a contribution
to the component $\pp_4$ of the magnetic charge vector $P$. A careful
calculation shows that the net additional contribution to $\pp_4$ 
due to this coupling is given by
\be \label{eq4}
\Delta_3 \pp_4 =
\qq_3  + \wh Q\cdot \wh P\, .
\ee
In this expression the contribution proportional to $\qq_3$ comes from
taking $F$ in \refb{eq3} to be the field strength associated with the
RR 2-form field of IIB, whereas the term proportional to $\wh Q\cdot
\wh P$ arises from taking $F$ to be the field strength associated with
the components of the RR 4-form field along various 2-cycles of $K3$.
\item Eqs.\refb{emag} and \refb{emag2} show that we have a net
D1-brane charge along $S^1$ equal to
\be \label{eq5}
\pp_3 = Q_1 - 1 -\wh P^2 / 2\, .
\ee
If we denote by $C^{(2)}$ the 2-form field of the original ten
dimensional type IIB string theory, then the effect of this charge is
to produce a background of the form:
\be \label{eq6}
dC^{(2)} \propto (Q_1 - 1 - \wh P^2/2) \, r^{-2}\, dr \wedge dt\wedge
dy\, .
\ee
Again one can verify explicitly that the right hand side of \refb{eq6}
is both closed and co-closed in the Taub-NUT background. We also have a 
component 
\be \label{eq7}
dC^{(2)} \propto \qq_5 \, dy\wedge \omega \, ,
\ee 
coming from the excitation of the collective coordinate of the
Kaluza-Klein monopole. This gives a source term for
$g^{\psi t}$ via the coupling proportional to
\be \label{eq8}
-\int \sqrt{-\det g} g^{\psi t} F_{\psi ry} F_{t}^{~ry} 
\ee
producing an additional contribution to the charge $\pp_4$
of the form
\be \label{eq9}
\Delta_4 \pp_4 = 
\qq_5(Q_1 - 1 - \wh P^2/2) \, .
\ee
\end{enumerate}

So far in our analysis we have taken into account possible additional
sources produced by the terms quadratic in the fields. What about
higher order terms? It is straightforward to show that the possible
effect of the higher order terms on the shift in the charges will
involve one or more powers of the type IIB string coupling. Since the
shift in the charges must be quantized, they cannot depend on
continuous moduli. Thus at least in the weakly coupled type IIB string
theory there are no additional corrections to the charges.
Incidentally, the same argument can be used to show that the shifts in
the charges must also be independent of the other moduli; thus it is
in principle sufficient to calculate these shifts at any particular
point in the moduli space.

Combining all the results we see that we have a net electric charge
vector $Q$ and a magnetic charge vector $P$ of the form:
\be \label{e1.6new}
Q = \pmatrix{\wh Q\cr \qq_1 \cr 0 \cr \qq_3 \cr -n + \qq_3 \qq_5 
+ \wh Q^2/2
\cr
\qq_5\cr -1}, \qquad P = \pmatrix{\wh P\cr \pp_1\cr 0 \cr Q_1
-1 -\wh P^2/2 \cr -J +\qq_3  + \wh Q\cdot \wh P
+  \qq_5(Q_1 - 1 - \wh P^2/2)\cr 1 \cr 0}
\, .
\ee
This has
\be \label{eq2new}
Q^2=2n, \qquad P^2 = 2(Q_1-1), \qquad Q\cdot P = J\, .
\ee
Thus the additional charges do not affect the relationship between the
invariants $Q^2$, $P^2$, $Q\cdot P$ and the original quantum numbers
$n$, $Q_1$ and $J$.

\sectiono{Dyon Spectrum} \label{s6}

Let us now turn to the analysis of the dyon spectrum in the presence
of these charges. For this we recall that in \cite{0605210} the dyon
spectrum was computed from three mutually non-interacting parts, --
the dynamics of the Kaluza-Klein monopole, the overall motion of the
D1-D5 system in the background of the Kaluza-Klein monopole and the
motion of the D1-branes relative to the D5-brane.  The precise
dynamics of the D1-branes relative to the D5-brane is affected by the
presence of the gauge field flux on the D5-brane since it changes the
non-commutativity parameter describing the dynamics of the gauge field
on the D5-brane world-volume\cite{9908142}.  As a result the moduli
space of D1-branes, described as non-commutative instantons in this
gauge theory\cite{9802068}, gets deformed.  However, we do not expect
this to change the elliptic genus of the corresponding conformal field
theory\cite{9608096} that enters the degeneracy formula.  With the
exception of the zero mode associated with the D1-D5 center of mass
motion in the Kaluza-Klein monopole background, the rest of the
contribution to the degeneracy came from the excitations involving
non-zero mode oscillators of the collective coordinates of the
Kaluza-Klein monopole and the collective coordinates associated with
the overall motion of the D1-D5 system\cite{0605210}.  This is not
affected either by switching on gauge field fluxes on the D5-brane
world-volume or the momenta or winding number of the collective
coordinates of the Kaluza-Klein monopole. On the other hand the
dynamics of the D1-D5-brane center of mass motion in the background
geometry is also not expected to be modified in the weakly coupled
type IIB string theory since in this limit the additional background
fields due to the additional charges are small compared to the one due
to the Kaluza-Klein monopole.  (For this it is important that the
additional charges do not involve any other Kaluza-Klein monopole or
NS 5-brane charge.) Thus we expect the degeneracy to be given by the
same function $f(n,Q_1,J)$ that appeared in the absence of the
additional charges. Using \refb{eq2new} we can now write
\be \label{efin}
d(Q,P) = f\left( {1\over 2} Q^2, {1\over 2}P^2+1, Q\cdot P\right)\, .
\ee
This is a generalization of \refb{e1.4} and shows that for the charge
vectors given in \refb{e1.6new}, the degeneracy $d(Q,P)$ depends on
the charges only through the combination $Q^2$, $P^2$ and $Q\cdot P$.

As was discussed in \cite{0702141}, the formula for the degeneracy for
a given charge vector can change across walls of marginal stability in
the moduli space.  Hence a given formula for the degeneracy makes
sense only if we specify how the region of the moduli space in which
we are carrying out our analysis is situated with respect to the walls
of marginal stability.  In the theory under consideration the moduli
space is the coset $\left(SL(2,\ZZZ)\backslash SL(2,\RRR)/U(1)\right)
\times \left(SO(6,22;\ZZZ)\backslash SO(6,22;\RRR)/SO(6)\times SO(22)
\right)$, parametrized by a complex modulus $\tau$ and a $28\times 28$
symmetric $SO(6,22)$ matrix $M$. For fixed $M$, the walls of marginal
stability are either straight lines in the $\tau$ plane, intersecting
the real axis at an integer, or circles intersecting the real axis at
rational points $a/c$ and $b/d$ with $ad-bc=1$, $a,b,c,d\in\ZZZ$.  The
precise shape of the circles and the slopes of the straight lines
depend on the modulus $M$ and the charge vector of the state under
consideration.  It was shown in \cite{0702141} that for the charge
vector given in \refb{e1.2} the region where the type IIB string
coupling and the angle between $S^1$ and $\wt S^1$ are small and the
other moduli are of order 1 can fall into one of the two domains in
the upper half $\tau$ plane. The first of these domains is bounded by
a pair of straight lines in the $\tau$ plane, passing through the
points 0 and 1 respectively, and a circle passing through the points 0
and 1. The second domain is bounded by a pair of straight lines
passing through the points $-1$ and 0 respectively and a circle
passing through the points $-1$ and 0.  Carrying out a similar
analysis for the modified charge vector \refb{e1.6new} one finds that
as long as all the charges are finite, the region of moduli space
where type IIB coupling is small falls inside the same domains, \i.e.\
domains bounded by a set of walls of marginal stability which
intersect the real $\tau$ axis at the same points.  This is just as
well; had the new charge vector landed us into a different domain in
the $\tau$ plane, our result \refb{efin} would be in contradiction
with the result of \cite{0702141} that in different domains bounded by
different walls of marginal stability the degeneracies are given by
different functions of $P^2$, $Q^2$ and $Q\cdot P$.

\sectiono{More General Charge Vector} \label{s9}

The charge vector given in \refb{e1.6new}, while more general than the
one considered in \cite{0605210}, is still not the most general charge
vector. Is it possible to extend our analysis to include more general
charge vectors? First of all note that $\pp_5$, representing the
number of D5-branes wrapped on $K3\times S^1$, was chosen to be an
arbitrary integer instead of 1 in \cite{0605210}. Thus we can
certainly take as our starting point the more general charge vector
where $Q_5$ in \refb{e1.2} is chosen to be an arbitrary integer
instead of 1. Our analysis up to \refb{eq2new} proceeds in a
straightforward manner (with $Q_1-1$ replaced by $Q_5(Q_1-Q_5)$).  The
issue, however, is how the additional charges affect the dyon
spectrum. In particular one needs to examine carefully the effect of
the gauge field flux on the D5-brane on the dynamics of the D1-D5
system, generalizing the analysis given in \cite{9608096}. However, as
long as, we do not switch on gauge field flux on the D5-branes, \i.e.\
consider configurations with $\wh P=0$, $l_1=0$, there is no
additional complication and the final degeneracy will still be given
by \refb{efin}. On the other hand following the analysis of
\cite{0702141} one can show that the region of the moduli space where
the type IIB string coupling and the angle between $S^1$ and $\wt S^1$
are small is still bounded by the same set $\BB_R$, $\BB_L$ of walls
of marginal stability.

In \refb{e1.6new} we have set the component $\qq_2$ of the electric
charge vector to zero even though we could switch it on by switching
on an NSNS sector 3-form field strength of the form $dy\wedge \omega$.
The reason for this was that this charge represents the number of NS
5-branes wrapped along $K3\times \wt S^1$ and the presence of NS
5-branes could have large backreaction on the geometry thereby
invalidating our analysis. We can, however, keep its effect small
compared to that of the original background produced by the
Kaluza-Klein monopole by taking the radius $R$ of $S^1$ to be large
compared to $\sqrt{\alpha'}$.  Since in the string metric the mass of
the Kaluza-Klein monopole is proportional to $R$ while the NS 5-brane
wrapped along $\wt S^1$ does not have such a factor, we can expect
that for large $R$ the effect of the background produced by the NS
5-brane will be small compared to that of the Kaluza-Klein monopole.
We can then analyze the system in the same manner as for the other
charges and conclude that the formula for the degeneracy in the
presence of this additional charge is still given by \refb{eq2new}.
One also finds that the region of the moduli space where the type IIB
string coupling and the angle between $S^1$ and $\wt S^1$ are small is
still bounded by the same set $\BB_R$, $\BB_L$ of walls of marginal
stability.

Let us now turn to $\qq_6$ which has been set equal to $-1$ in 
\refb{e1.6new}. This is the number of Kaluza-Klein monopoles
associated with the compactification circle $\wt S^1$. Changing
this number would require us to study the dynamics of
multiple Kaluza-Klein monopoles. While, in principle, this can be
done, this will certainly require a major revision of the analysis
done so far. Thus there does not seem to be a minor variation of
our analysis that can change the charge $\qq_6$ to 
any other integer.

This leaves us with the components $\pp_2$ and $\pp_6$ both of which
have been set to 0 in \refb{e1.6new}. $\pp_2$ represents the number of
NS 5-branes wrapped on $S^1$. Switching this charge on would require
us to introduce explicit NS 5-brane background and study the dynamics
of D-branes in such a background. This would require techniques quite
different from the one used so far.  On the other hand, the component
$\pp_6$ represents the Kaluza-Klein monopole charge associated with
the compact circle $S^1$. This also causes significant change in the
background geometry and calculation of the spectrum of such
configurations would require fresh analysis.

\bigskip

\noindent {\bf Acknowledgement:} We would like to thank Justin David for 
many
useful discussions and collaboration at early stages of this work.

\appendix

\sectiono{Normalization and Sign Conventions} \label{sa}

In this appendix we shall describe the various normalization and sign
conventions we use during our analysis.  We begin by describing the
ten dimensional action of type IIB string theory that appears in the
first description:
\ben \label{et.1}
S &=& {1\over (2\pi)^7 (\alpha')^4} \int d^{10} x \, \sqrt{-\det g}
\Bigg[ e^{-2\Phi} \Bigg( R + 4 \p_M \Phi \p^M\Phi 
 - {1\over 2\cdot
3!} H_{MNP}H^{MNP} \Bigg) \nonumber \\
&& 
-{1\over 2} F^{(1)}_M F^{(1)M}  -{1\over 2 \cdot 3!}
\wt F^{(3)}_{MNP} \wt F^{(3)MNP}  
-{1\over 4 \cdot 5!} \wt F^{(5)}_{M_1\cdots M_5}
\wt F^{(5)M_1\cdots M_5}\Bigg] \nonumber \\
&&
+ {1\over 2(2\pi)^7 (\alpha')^4} \int C^{(4)} \wedge \wt F^{(3)}
\wedge H \, ,
\een
where 
\ben \label{et.2} 
H= d B, \quad F^{(1)} = dC^{(0)}, \quad F^{(3)} = dC^{(2)}, \quad
\,  F^{(5)} = d C^{(4)} \nonumber \\
\wt F^{(3)} = F^{(3)} - C^{(0)} H, \quad \wt 
F^{(5)} = F^{(5)} - {1\over 
2} C^{(2)} \wedge H
+ {1\over 2} B \wedge F^{(3)}\, ,
\een
$g_{MN}$ denotes the string metric, $B_{MN}$ denotes the NSNS 2-form
fields, $\Phi$ denotes the dilaton and $C^{(k)}$ denotes the RR
$k$-form field.  The field strengths $dC^{(k)}$ are subject to the
relations $* d C^{(k)} = (-1)^{k(k-1)/2} dC^{(8-k)} +\cdots$ where $*$
denotes Hodge dual taken with respect to the string metric and
$\cdots$ denote terms quadratic and higher order in the fields.  For
$k=4$ this gives a constraint on $C^{(4)}$ whereas for $k> 4$ this
defines the field $C^{(k)}$.  In computing the Hodge dual in the first
description we shall use the convention that on $S^1\times \wt
S^1\times \RRR^{3,1}$ we have $\epsilon^{ty\psi r\theta\phi}>0$ where
$r$, $\theta$, $\phi$ and $t$ denote the spherical polar coordinates
and the time coordinate of the (3+1) dimensional non-compact
space-time and $y$ and $\psi$ denote coordinates of $S^1$ and $\wt
S^1$ respectively, each normalized to have period $2\pi
\sqrt{\alpha'}$. Inside $K3$ we use the standard volume form on K3 to
define the $\epsilon$ tensor. Our normalization conventions are
consistent with that of \cite{9910053}.

As is well known, the moduli space of $K3$ with NSNS 2-form fields
switched on, is labelled by elements of the coset
$SO(4,20;\ZZZ)\backslash SO(4,20)/(SO(4)\times SO(20))$.  These
elements may be parametrized by a symmetric SO(4,20) matrix $\wt M$
and we choose the coordinate system on this coset in such a way that
the identity matrix represents a $K3$ of volume
$(2\pi\sqrt{\alpha'})^4$ in string metric, with the NSNS 2-form fields
set to zero.

The low energy effective action of heterotic string theory on $T^4$
that appears in the second description has the form:
\ben \label{et.3}
&& {1\over (2\pi)^3 (\alpha')^2} \, \int d^6 x\, \sqrt{-\det g}
\, e^{-2\Phi} \,
\Bigg[R + 4 \p_\alpha \Phi \p^\alpha\Phi
 - {1\over 2\cdot
3!} H_{\alpha\beta\gamma}
H^{\alpha\beta\gamma} -{1\over 8} Tr 
(\p_\alpha \wt M \wt L \p^\alpha \wt M \wt 
L) \nonumber \\
&& \qquad - \FF^{(a)}_{\alpha\beta} (\wt L \wt M \wt L)_{ab} 
\FF^{(b)\alpha\beta}\Bigg] 
\een
where $\wt L$ is a fixed $24\times 24$ matrix with 4 positive and 20
negative eigenvalues, $\wt M$ is a $24\times 24$ symmetric matrix
valued scalar field satisfying $\wt M \wt L \wt M=\wt L$, and
$\FF^{(a)}_{\alpha\beta}$ for $1\le a\le 24$, $0\le\alpha,\beta\le 5$
are the field strengths associated with 24 U(1) gauge fields
$\AAA_\alpha^{(a)}$ obtained by heterotic string compactification on
$T^4$.  The fields $g_{\alpha\beta}$, $B_{\alpha\beta}$ and $\Phi$ are
the string metric, NSNS 2-form field and the six dimensional dilaton
of the heterotic string theory and should be distinguished from those
appearing in \refb{et.1}.  Upon further compactification on $\wh
S^1\times S^1$ labelled by $x^4\equiv \chi$ and $x^5\equiv y$, both
normalized to have period $2\pi\sqrt{\alpha'}$, we get four more gauge
fields $A_\mu^{(i)}$ ($1\le i\le 4$, $0\le\mu,\nu\le 3$) and a
$4\times 4$ symmetric matrix valued scalar field $\bar M$ defined via
the relations:
\ben \label{etenfour}
&& \wh G_{mn} \equiv g_{mn}, \quad
\wh B_{mn} \equiv B_{mn}\, ,
\qquad
m,n=4,5\, ,
\nonumber  \\
&& \bar M =
\pmatrix{ \wh G^{-1} & \wh G^{-1} \wh B \cr -\wh B \wh G^{-1} & \wh
G - \wh B \wh G^{-1} \wh B} \nonumber \\
&& A^{(m-3)}_\mu = {1\over 2} (\wh G^{-1})^{mn}
G^{(10)}_{m\mu} , \quad
A^{(m-1)}_\mu = {1\over 2} B^{(10)}_{m\mu} -
\wh B_{mn} A^{(m-3)}_\mu, \nonumber \\
&& \qquad  4\le m,n\le 5, \quad
0\le \mu, \nu \le 3 \, .
\een
For simplicity we have set the Wilson lines of the gauge fields 
$\AAA_\alpha^{(a)}$ along $S^1$ and $\wh S^1$ to zero.  In the 
$\alpha'=16$ unit the electric and magnetic charges 
$(k_3,\cdots k_6, l_3,\cdots l_6)$ appearing in eq.\refb{e1.1} are 
related to the asymptotic values of the gauge field strengths 
$F_{\mu\nu}^{(i)}=\p_{\mu}A^{(i)}_\nu - \p_{\nu}A^{(i)}_\mu$ via 
the relations\cite{0508042}
\be \label{et.5}
(\bar L\bar M \bar L )_{ij} F^{(j)}_{rt}\bigg|_\infty =  
{ k_{i+2}\over r^2}\, ,
\qquad \bar 
L_{ij} \, F^{(j)}_{\theta\phi}\bigg|_\infty =  l_{i+2} \sin\theta\, , 
\qquad \bar L \equiv \pmatrix{ 0_2 & I_2\cr 
I_2 & 
0_2}\, .
\ee
The other charges $\wh Q$, $k_1$, $k_2$ and $\wh P$, $l_1$, $l_2$ 
appearing in \refb{e1.1} can be related to the asymptotic values of the 
gauge field strengths $\FF^{(a)}_{rt}$ and $\FF^{(a)}_{\theta\phi}$
in a similar manner. 

The chain of duality transformations taking us from the first to the
second description are chosen so that at the linearized level the
first S-duality transformation of IIB acts as $C^{(2)}\to B$, $B\to
-C^{(2)}$, and the next $R\to 1/R$ duality transformations of $\wt
S^1$ acts as $g_{\psi\mu}\to - B_{\chi\mu}$, $B_{\psi\mu}\to -
g_{\chi\mu}$ together with appropriate transformations on the various
RR gauge fields.  The final string string duality transformation acts
via a Hodge duality transformation in six dimensions on the NS sector
3-form field strength with $\epsilon^{t\chi y r\theta\phi}>0$, and
maps various four dimensional gauge fields arising from various
components of the RR sector fields to the 24 gauge fields in heterotic
string theory on $T^4$.

Finally, we use the following convention for the signs of the charges
carried by various branes in the first description.  If $F^{(3)}\equiv
d C^{(2)}$ denotes the RR 3-form field strength, then asymptotically a
D1-brane along $S^1$ will carry positive $F^{(3)}_{yrt}$, a D5-brane
along $\wt S^1\times K3$ will carry positive $F^{(3)}_{\theta y\phi}$,
a D1-brane along $\wt S^1$ will carry positive $F^{(3)}_{\psi rt}$ and
a D5-brane along $S^1\times K3$ will carry negative $F^{(3)}_{\theta
  \psi\phi}$.  The same convention is followed for fundamental string
and NS 5-brane with $F^{(3)}$ replaced by the NSNS 3-form field
strength $H=dB$.  A state carrying positive momentum along $S^1$ or
$\wt S^1$ is defined to be the one which produces positive $\p_r
g_{yt}$ or $\p_r g_{\psi t}$, and a positively charged Kaluza-Klein
monopole associated with the circle $S^1$ or $\wt S^1$ is defined to
be the one that carries positive $\p_\theta g_{y\phi}$ or $\p_\theta
g_{\psi\phi}$ asymptotically.  Note that in this convention the
asymptotic configuration for $F^{(7)}\equiv dC^{(6)}$ around a
D5-brane wrapped on $K3\times S^1$ or $K3\times \wt S^1$ will have
negative $F^{(7)}_{(K3)yrt}$ or $F^{(7)}_{(K3)\psi r t}$, with the
subscript $_{(K3)}$ denoting components of $F^{(7)}$ along the volume
form of $K3$.

The same conventions are followed for the signs of the charges carried
by various states in the second description, with the coordinate
$\psi$ of $\wt S^1$ replaced by the coordinate $\chi$ of $\wh S^1$.


\baselineskip 12.2pt


\begin{thebibliography}{99}

\small 
\parskip=-2pt
  
\bibitem{9607026}
R.~Dijkgraaf, E.~P.~Verlinde and H.~L.~Verlinde,
``Counting dyons in N = 4 string theory,''
Nucl.\ Phys.\ B {\bf 484}, 543 (1997)
[arXiv:hep-th/9607026].

\bibitem{0505094}
D.~Shih, A.~Strominger and X.~Yin,
``Recounting dyons in N = 4 string theory,''
arXiv:hep-th/0505094.

\bibitem{0506249}
D.~Gaiotto,
``Re-recounting dyons in N = 4 string theory,''
arXiv:hep-th/0506249.

\bibitem{0510147}
  D.~P.~Jatkar and A.~Sen,
  ``Dyon spectrum in CHL models,''
  JHEP {\bf 0604}, 018 (2006)
  [arXiv:hep-th/0510147].

\bibitem{0602254}
  J.~R.~David, D.~P.~Jatkar and A.~Sen,
  ``Product representation of dyon partition function in CHL models,''
  JHEP {\bf 0606}, 064 (2006)
  [arXiv:hep-th/0602254].
  
\bibitem{0603066}
  A.~Dabholkar and S.~Nampuri,  
  ``Spectrum of dyons and black holes in 
  CHL orbifolds using Borcherds lift,''
  arXiv:hep-th/0603066.

\bibitem{0605210}
  J.~R.~David and A.~Sen,
  ``CHL dyons and statistical entropy function from D1-D5 system,''
  JHEP {\bf 0611}, 072 (2006)
  [arXiv:hep-th/0605210].

\bibitem{0607155}
  J.~R.~David, D.~P.~Jatkar and A.~Sen,
  ``Dyon spectrum in N = 4 supersymmetric type II string theories,''
  arXiv:hep-th/0607155.


\bibitem{0609109}
  J.~R.~David, D.~P.~Jatkar and A.~Sen,
  ``Dyon spectrum in generic N = 4 supersymmetric Z(N) orbifolds,''
  arXiv:hep-th/0609109.

\bibitem{0612011}
  A.~Dabholkar and D.~Gaiotto,
  ``Spectrum of CHL dyons from genus-two partition function,''
  arXiv:hep-th/0612011.

\bibitem{0702141}
  A.~Sen,
  ``Walls of marginal stability and dyon spectrum in N = 4 supersymmetric
  string theories,''
  arXiv:hep-th/0702141.
  
\bibitem{0702150}
  A.~Dabholkar, D.~Gaiotto and S.~Nampuri,
  ``Comments on the spectrum of CHL dyons,''
  arXiv:hep-th/0702150.
  
\bibitem{0508042}
  A.~Sen,
  ``Entropy function for heterotic black holes,''
  JHEP {\bf 0603}, 008 (2006)
  [arXiv:hep-th/0508042].

\bibitem{9910053}
  R.~C.~Myers,
  ``Dielectric-branes,''
  JHEP {\bf 9912}, 022 (1999)
  [arXiv:hep-th/9910053].
  
\bibitem{brill}
D.~Brill, Phys. Rev. B133 (1964) 845.

  \bibitem{pope}
 C.~N.~Pope,
  ``Axial Vector Anomalies And The Index Theorem In Charged 
 Schwarzschild And
  Taub - Nut Spaces,''
  Nucl.\ Phys.\ B {\bf 141}, 432 (1978).

\bibitem{9705212}
  A.~Sen,
  ``Kaluza-Klein dyons in string theory,''
  Phys.\ Rev.\ Lett.\  {\bf 79}, 1619 (1997)
  [arXiv:hep-th/9705212].


\bibitem{9908142}
  N.~Seiberg and E.~Witten,
  ``String theory and noncommutative geometry,''
  JHEP {\bf 9909}, 032 (1999)
  [arXiv:hep-th/9908142].
  
  \bibitem{9802068}
  N.~Nekrasov and A.~S.~Schwarz,
  ``Instantons on noncommutative 
  R**4 and (2,0) superconformal six  dimensional
  theory,''
  Commun.\ Math.\ Phys.\  {\bf 198}, 689 (1998)
  [arXiv:hep-th/9802068].
  

  
\bibitem{9608096}
  R.~Dijkgraaf, G.~W.~Moore, E.~P.~Verlinde and H.~L.~Verlinde,
  ``Elliptic genera of symmetric products and second quantized strings,''
  Commun.\ Math.\ Phys.\  {\bf 185}, 197 (1997)
  [arXiv:hep-th/9608096].

  

\end{thebibliography}
\end{document}